
\input harvmac

\Title{\vbox{\baselineskip12pt
\hbox{CTP-TAMU-50/94}
\hbox{CERN-TH/95-122}
\hbox{UCLA 95/TEP-15}
\hbox{McGill/95-23}
\hbox{NI-94-035}
\hbox{hep-th/9506057}
}}
{\vbox{\centerline {Supersymmetry and Dual String Solitons}}}
\centerline{M.~J.~Duff$^1$\footnote{$^*$}{Supported in part
by NSF
grant PHY-9411543.}, Sergio Ferrara$^{2,3}$
\footnote{$^\dagger$}{Supported in part by DOE grants
DE-AC0381-ER50050
and DOE-AT03-88ER40384, Task E and by EEC Science Program
SC1$^*$CI92-0789.}, Ramzi R.~Khuri$^{2,4}$ and Joachim
Rahmfeld$^1$}
\bigskip\centerline{$^1$
{\it Physics Department, Texas A\&M University,
 College Station, TX 77843-4242 USA}}
\bigskip\centerline{$^2${\it CERN,
 CH-1211, Geneva 23, Switzerland}}
\bigskip\centerline{$^3${\it Physics Department,
University of California, Los Angeles, CA 90024-1547 USA}}
\bigskip\centerline{$^4$
{\it Physics Department, McGill University, Montreal, PQ,
H3A 2T8 Canada}}
\vskip .3in
We present new classes of string-like soliton
solutions in ($N=1$; $D=10$), ($N=2$; $D=6$)
 and ($N=4$; $D=4$) heterotic string theory.
Connections are made between the solution-generating
subgroup of the $T$-duality group of
the compactification and the number of spacetime supersymmetries
broken. Analogous solutions are also noted in ($N=1,2$; $D=4$)
compactifications, where a different form of supersymmetry
breaking arises.

\vskip .3in
\Date{\vbox{\baselineskip12pt
\hbox{CERN-TH/95-122}
\hbox{June 1995}}}

\def\sqr#1#2{{\vbox{\hrule height.#2pt\hbox{\vrule width
.#2pt height#1pt \kern#1pt\vrule width.#2pt}\hrule height.#2pt}}}
\def\Box{\mathchoice\sqr64\sqr64\sqr{4.2}3\sqr33}

\lref\prep{M. J. Duff, R. R. Khuri and J. X. Lu,
 NI-94-017, CTP/TAMU-67/92,
McGill/94/53, CERN-TH.7542/94, hepth/9412184
(to appear in Physics Reports).}

\lref\fabjr{M. Fabbrichesi, R. Jengo and K. Roland, Nucl.
Phys. {\bf B402} (1993) 360.}

\lref\jxthesis{J. X. Lu, {\it Supersymmetric Extended
Objects}, Ph.D. Thesis, Texas A\&M University (1992),
UMI {\bf 53 08B}, Feb. 1993.}

\lref\coleman{S. Coleman, ``Classical Lumps and Their Quantum
Descendants'', in {\it New Phenomena in Subnuclear Physics},
ed A. Zichichi (Plenum, New York, 1976).}

\lref\jackiw{R. Jackiw, Rev. Mod. Phys. {\bf 49} (1977) 681.}

\lref\hink{M. B. Hindmarsh and T. W. B. Kibble, SUSX-TP-94-74,
IMPERIAL/TP/94-95/5, NI 94025, hepph/9411342.}

\lref\senmod{A. Sen, Mod. Phys. Lett. {\bf A8} (1993) 2023.}

\lref\kikyam{K. Kikkawa and M. Yamasaki, Phys. Lett. {\bf B149}
(1984) 357.}

\lref\sakai{N. Sakai and I. Senda, Prog. Theor. Phys. {\bf 75}
(1986) 692.}

\lref\bush{T. Busher, Phys. Lett. {\bf B159} (1985) 127.}

\lref\nair{V. Nair, A. Shapere, A. Strominger and F. Wilczek,
Nucl. Phys. {\bf B322} (1989) 167.}

\lref\duff{M. J. Duff, Nucl. Phys. {\bf B335} (1990)
610.}

\lref\tseyv{A. A. Tseytlin and C. Vafa, Nucl. Phys. {\bf B372}
(1992) 443,}

\lref\tseycqg{A. A. Tseytlin, Class. Quantum Grav. {\bf 9}
(1992) 979.}

\lref\givpr{A. Giveon, M. Porrati and E. Rabinovici,
(to appear in Phys. Rep. C).}

\lref\gibkal{G. W. Gibbons and R. Kallosh, NI-94003,
hepth/9407118.}

\lref\vafw{C. Vafa and E. Witten, Nucl. Phys. {\bf B431}
(1994) 3.}

\lref\seiw{N. Seiberg and E. Witten, Nucl. Phys. {\bf B426}
(1994) 19.}

\lref\seiwone{N. Seiberg and E. Witten, Nucl. Phys. {\bf B431}
(1994) 484.}

\lref\cerdfv{A. Ceresole, R. D'Auria, S. Ferrara and
A. Van Proeyen, CERN-TH 7510/94, POLFIS-TH.08/94,
UCLA 94/TEP/45, KUL-TF-94/44, hepth/9412200.}

\lref\cerdfvone{A. Ceresole, R. D'Auria, S. Ferrara and
A. Van Proeyen, CERN-TH 7547/94, POLFIS-TH.01/95,
UCLA 94/TEP/45, KUL-TF-95/4, hepth/9502072.}

\lref\gauh{J. Gauntlett and J. H. Harvey, EFI-94-30,
hepth/9407111.}

\lref\frak{P. H. Frampton and T. W. Kephart, IFP-708-UNC,
VAND-TH-94-15.}

\lref\hawhr{S. W. Hawking, G. T. Horowitz and S. F. Ross,
NI-94-012, DAMTP/R 94-26, UCSBTH-94-25, gr-qc/9409013.}

\lref\ellmn{J. Ellis, N. E. Mavromatos and D. V. Nanopoulos,
Phys. Lett. {\bf B278} (1992) 246.}

\lref\kaln{S. Kalara and N. Nanopoulos, Phys. Lett. {\bf B267}
(1992) 343.}

\lref\glasgow{M. J. Duff, hepth/9410210, in Proceedings of the
   High Energy Physics Conference, Glasgow June 1994.}

\lref\dufnew{M. J. Duff, NI-94-033, CTP-TAMU-49/94,
hepth/9501030.}

\lref\bhs{R. R. Khuri, Helv. Phys. Acta {\bf 67} (1994) 884.}

\lref\cecfg{S. Cecotti, S. Ferrara and L. Girardello,
Int. J. Mod. Phys. {\bf A4} (1989) 2475.}

\lref\ferquat{S. Ferrara and S. Sabharwal,
Nucl. Phys. {\bf B332} (1990) 317.}

\lref\fere{R. C. Ferrell and D. M. Eardley,
Phys. Rev. Lett. {\bf 59} (1987) 1617.}

\lref\reyt{S. J. Rey and T. R. Taylor, Phys. Rev. Lett. {\bf 71}
(1993) 1132.}

\lref\senzwione{A. Sen and B. Zwiebach, Nucl. Phys.
{\bf B414} (1994) 649.}

\lref\senzwitwo{A. Sen and B. Zwiebach, Nucl. Phys.
{\bf B423} (1994) 580.}

\lref\bddo{T. Banks, A. Dabholkar, M. R. Douglas and
M. O'Loughlin, Phys. Rev. {\bf D45} (1992) 3607.}

\lref\bos{T. Banks, M. O'Loughlin and A. Strominger,
Phys. Rev. {\bf D47} (1993) 4476.}

\lref\kir{E. Kiritsis, Nucl. Phys. {\bf B405} (1993) 109.}

\lref\dufr{M. J. Duff and J. Rahmfeld, Phys. Lett. B
{\bf 345} (1995) 441.}

\lref\duffkk{M. J. Duff, NI-94-015, CTP-TAMU-22/94,
hepth/9410046.}

\lref\bakone{I. Bakas, Nucl. Phys. {\bf B428} (1994) 374.}

\lref\baktwo{I. Bakas, Phys. Lett. {\bf B343} (1995) 103.}

\lref\bakthree{I. Bakas, Phys. Lett. {\bf B349} (1995) 448.}

\lref\baksfet{I. Bakas and K. Sfetsos, CERN-TH-95-16,
hepth/9502065.}

\lref\back{C. Bachas and E. Kiritsis, Phys. Lett. {\bf B325}
(1994) 103.}

\lref\bko{E. Bergshoeff, R. Kallosh and T. Ortin,
UG-8/94, SU-ITP-94-19, QMW-PH-94-13, hepth/9410230.}

\lref\vilsh{A. Vilenkin and E. P. Shellard,
{\it Cosmic String and Other Topological Defects},
(Cambridge University Press, 1994).}

\lref\bfrm{M. Bianchi, F. Fucito, G. C. Rossi and M. Martellini,
hepth/9409037.}

\lref\hult{C. M. Hull and P. K. Townsend, QMW-94-30,
R/94/33, hepth/9410167.}

\lref\ght{G. W. Gibbons, G. T. Horowitz and P. K. Townsend,
R/94/28, UCSBTH-94-35, hepth/9410073.}

\lref\dufkmr{M. J. Duff, R. R. Khuri, R. Minasian and
J. Rahmfeld, Nucl. Phys. {\bf B418} (1994) 195.}

\lref\sentd{A. Sen, Nucl. Phys. {\bf B434} (1995) 179.}

\lref\maha{J. Maharana, NI-94023, hepth/9412235.}

\lref\gresss{M. B. Green and J. Schwarz, Phys. Lett. {\bf B136}
(1984) 367.}

\lref\sie{W. Siegel, Phys. Lett. {\bf B128} (1983) 397.}

\lref\dir {P. A. M. Dirac, Pro. R. Soc. {\bf A133} (1931) 60.}

\lref\tho {G. t'Hooft, Nucl. Phys. {\bf B79} (1974) 276.}

\lref\pol {A. M. Polyakov, Sov. Phys. JETP Lett. {\bf 20}
(1974) 194.}

\lref\mono {C. Montonen and D. Olive, Phys. Lett. {\bf B72}
(1977) 117.}

\lref\col {S. Coleman, Phys. Rev. {\bf D11} (1975) 2088.}

\lref\grohmr {D. J. Gross, J. A. Harvey, E. Martinec and
 R. Rohm,
Nucl. Phys. {\bf B256} (1985) 253.}

\lref\ginone {P. Ginsparg,
 {\it Conformal Field Theory}, Lectures given
at Trieste Summer School, Trieste, Italy, 1991.}

\lref\calhstwo {C. Callan, J. Harvey and A. Strominger,
 {\it Supersymmetric String Solitons}, Lectures given
at Trieste Summer School, Trieste, Italy, 1991.}

\lref\sch {J. H. Schwarz, {\it Supersymmetry and Its
 Applications}
ed G. W. Gibbons {\it et al} (Cambridge University Press,
1986).}

\lref\huglp {J. Hughes, J. Liu and J. Polchinski, Phys. Lett.
{\bf B180} (1986)
370.}

\lref\berst {E. Bergshoeff, E. Sezgin and P. K. Townsend,
 Phys. Lett. {\bf B189} (1987) 75.}

\lref\achetw {A. Achucarro, J. Evans, P. K.  Townsend and
D. Wiltshire,
 Phys.
Lett. {\bf B198} (1987) 441.}

\lref\belpst {A. A. Belavin, A. M. Polyakov, A. S. Schwartz and
Yu. S. Tyupkin,
Phys. Lett. {\bf B59} (1975) 85.}

\lref\oset {D. O'Se and D. H. Tchrakian, Lett. Math. Phys.
{\bf 13} (1987) 211.}

\lref\groks {B. Grossman, T. W. Kephart and J. D. Stasheff,
 Commun. Math. Phys. {\bf 96} (1984) 431;
Commun. Math. Phys. {\bf 100} (1985) 311.}

\lref\grokstwo {B. Grossman, T. W. Kephart and J. D. Stasheff,
 Phys. Lett. {\bf B220}
 (1989)  431.}

\lref\tch{D. H. Tchrakian, Phys. Lett. {\bf B150}  (1985)  360.}

\lref\fubn{S. Fubini and H. Nicolai,
 Phys. Lett. {\bf B155}  (1985)  369}

\lref\fain{D. B. Fairlie and J. Nuyts,
 J. Phys. {\bf A17}  (1984)  2867.}

\lref\str {A. Strominger,  Nucl. Phys. {\bf B343}
(1990) 167.}

\lref\duflhs {M. J. Duff and J. X. Lu,
Phys. Rev. Lett. {\bf 66} (1991) 1402.}

\lref\godo {P. Goddard and D. Olive, Rep. Prog. Phys. {\bf 41}
(1978) 1357.}

\lref\colone {S. Coleman, Proc. 1975 Int. School on Subnuclear
Physics,
Erice, ed A. Zichichi  (Plenum, New York, 1977); Proc. 1981
Int. School
on Subnuclear Physics, Erice, ed A. Zichichi (Plenum, New York,
1983).}

\lref\wuy {T. T. Wu and C. N. Yang, Nucl. Phys. {\bf B107}
(1976) 365.}

\lref\dufldl {M. J. Duff and J. X. Lu, Class. Quantum Grav.
{\bf 9} (1992) 1.}

\lref\tei {C. Teitelboim,
Phys. Lett. {\bf B167} (1986) 69.}

\lref\nep {R. I. Nepomechie,
Phys. Rev. {\bf D31} (1984) 1921.}

\lref\raj {R. Rajaraman, {\it Solitons and Instantons}
(North--Holland, Amsterdam, 1982).}

\lref\wito {E. Witten and D. Olive, Phys. Lett.
{\bf B78} (1978) 97.}

\lref\pras {M. K. Prasad and C. M. Sommerfield, Phys. Rev. Lett.
{\bf 35} (1975) 760.}

\lref\corg {E. Corrigan and P. Goddard, Commun. Math. Phys.
{\bf 80} (1981)
575.}

\lref\godno {P. Goddard, J. Nuyts and D. Olive, Nucl. Phys.
{\bf B125} (1977)
1.}

\lref\osb {H. Osborn, Phys. Lett. {\bf B83} (1979) 321.}

\lref\egugh {T. Eguchi, P. B. Gilkey and A. J. Hanson,
Phys. Rep. {\bf 66}
(1980) 213.}

\lref\corf {E. F. Corrigan and D. B. Fairlie, Phys. Lett.
{\bf B67} (1977) 69.
}

\lref\atidhm {M. F. Atiyah, V. G. Drinfeld, N. J. Hitchin and
Y. I. Manin,
Phys. Lett. {\bf A65} (1978) 185.}

\lref\dun {A. R. Dundarer,
 Mod. Phys. Lett. {\bf A5} (1991) 409.}

\lref\cha {A. H. Chamseddine, Phys. Rev. {\bf D24} (1981) 3065.}

\lref\berrwv {E. A. Bergshoeff, M. de Roo, B. de Wit and
 P. van Nieuwenhuizen,
Nucl. Phys. {\bf B195}  (1982)  97}

\lref\cham{G. F. Chapline and N. S. Manton, Phys. Lett.
{\bf B120} (1983) 105.}

\lref\gatn {S. J. Gates and H. Nishino, Phys. Lett. {\bf B173}
(1986) 52.}

\lref\sala{A. Salam and E. Sezgin, Physica Scripta {\bf 32}
(1985) 283.}

\lref\duf {M. J. Duff,  Class.
 Quantum  Grav. {\bf 5} (1988) 189.}

\lref\duflfb {M. J. Duff and J. X. Lu, Nucl. Phys. {\bf B354}
(1991) 141.}

\lref\dabghr {A. Dabholkar, G. W. Gibbons, J. A. Harvey and
F. Ruiz Ruiz,
 Nucl. Phys. {\bf B340} (1990) 33.}

\lref\berdps {E. Bergshoeff, M. J. Duff, C. N. Pope and
E. Sezgin,
 Phys. Lett.
{\bf B199} (1987) 69.}

\lref\tow {P. K. Townsend,
Phys. Lett. {\bf B202} (1988) 53.}

\lref\dufhis {M. J. Duff, P. S. Howe, T. Inami and K. Stelle,
 Phys. Lett.
{\bf B191} (1987) 70.}

\lref\dufs {M. J. Duff and K. Stelle, Phys. Lett. {\bf B253}
(1991) 113.}

\lref\berst {E. Bergshoeff, E. Sezgin and P. K. Townsend, Ann.
 Phys. {\bf 199}
(1990) 340.}

\lref\duflrsfd {M. J. Duff and J. X. Lu, Nucl. Phys. {\bf B354}
(1991) 129.}

\lref\gresone {M. Green and J. Schwarz, Phys. Lett. {\bf B151}
(1985) 21.}

\lref\bercgw {C. W. Bernard, N. H. Christ, A. H. Guth and
E. J. Weinberg,
 Phys. Rev.
 {\bf D16} (1977) 2967.}

\lref\hars {J. Harvey and A. Strominger,
 Phys. Rev. Lett. {\bf 66} (1991) 549.}

\lref\gres {M. Green and J. Schwarz, Phys. Lett. {\bf B149}
(1984) 117.}

\lref\elljm {J. Ellis, P. Jetzer and L. Mizrachi,
Nucl. Phys. {\bf B303} (1988)
1.}

\lref\dixds {J. Dixon, M. J. Duff and E. Sezgin, Phys. Lett.
{\bf B279} (1992) 265.}

\lref\berrs {E. Bergsheoff, M. Rakowski and E. Sezgin, Phys.
 Lett. {\bf
B185}  (1987)  371}

\lref\berd{E. Bergsheoff and M. de Roo, Nucl. Phys. {\bf B328}
(1989)
439}

\lref\dersw{M. de Roo, H. Suelmann and A. Wiedemann, preprint
UG--1/92  (1992).}

\lref\gresw {M. Green, J. Schwarz and E. Witten,
{\it Superstring
theory} (Cambridge University Press, 1987).}

\lref\duflloop {M. J. Duff and J. X. Lu, Nucl. Phys. {\bf B357}
  (1991)  534.}

\lref\ven {G. Veneziano,
Europhys. Lett. {\bf 2}  (1986)  199.}

\lref\cain {Y. Cai and C. A. Nunez, Nucl. Phys. {\bf B287}
(1987)  41}

\lref\gros{D. J. Gross and J. Sloan, Nucl. Phys. {\bf B291}
(1987)  41.}

\lref\ellm {J. Ellis and L. Mizrachi,
  Nucl. Phys. {\bf B327}  (1989)  595.}

\lref\calfmp {C. G. Callan, D. Friedan, E. J. Martinec and
M. J. Perry,
Nucl. Phys. {\bf B262}  (1985)  593.}

\lref\grestwo {M. Green and J. Schwarz, Phys. Lett. {\bf B173}
 (1986)  52.}

\lref\lin {U. Lindstrom, in Supermembranes and Physics in 2 + 1
Dimensions, ed. M. J. Duff, C. N. Pope and E. Sezgin
(World Scientific,
Singapore) (1990).}

\lref\dufone {M. J. Duff, Class. Quantum Grav. {\bf 6}  (1989)
1577.}

\lref\callny {C. Callan, C. Lovelace, C. Nappi and S. Yost,
Nucl. Phys.
{\bf B308}  (1988)  221.}

\lref\frat {E. Fradkin and A. Tseytlin, Phys. Lett. {\bf B158}
(1985)  316.}

\lref\duflselft {M. J. Duff and J. X. Lu, Phys. Lett.
{\bf B273}  (1991)  409.}

\lref\hors {G. Horowitz and A. Strominger, Nucl. Phys.
{\bf B360}  (1991) 197. }

\lref\witone {E. Witten,  Phys. Lett. {\bf B86}  (1979)
283.}

\lref\schone {J. Schwarz, Nucl. Phys. {\bf B226}  (1983)  269.}

\lref\zwa {D. Zwanziger, Phys. Rev. {\bf 176}  (1968)  1480,
1489.}

\lref\schwing {J. Schwinger, Phys. Rev. {\bf 144} (1966) 1087;
{\bf 173}
(1968) 1536.}

\lref\gibt{G.W. Gibbons and P.K. Townsend, Phys. Rev. Lett.
{\bf 71} (1993)
3754.}

\lref\duflblacks {M. J. Duff and J. X. Lu,
 Nucl. Phys. {\bf B416} (1994) 301.}

\lref\dufklsin {M. J. Duff, R. R. Khuri and J. X. Lu,
 Nucl. Phys. {\bf B377}
(1992) 281.}

\lref\calk {C.~G.~Callan and R.~ R.~Khuri,
Phys. Lett. {\bf B261} (1991) 363.}

\lref\gib {G. W. Gibbons, Nucl. Phys. {\bf B207} (1982) 337.}

\lref\gibm {G. W. Gibbons and K. Maeda, Nucl. Phys. {\bf B298}
(1988) 741.}

\lref\rey{S. J. Rey, in Proceedings of Tuscaloosa
Workshop on Particle Physics, (Tuscaloosa, Alabama, 1989).}

\lref\reyone {S. J.~Rey, Phys. Rev. {\bf D43} (1991) 526.}

\lref\antben {I.~Antoniadis, C.~Bachas, J.~Ellis and
 D.~V.~Nanopoulos,
Phys. Lett. {\bf B211} (1988) 393.}

\lref\antbenone {I.~Antoniadis, C.~Bachas, J.~Ellis and
D.~V.~Nanopoulos,
Nucl. Phys. {\bf B328} (1989) 117.}

\lref\mett {R.~R.~Metsaev and A.~A.~Tseytlin, Phys. Lett.
{\bf B191} (1987) 354.}

\lref\mettone {R.~R.~Metsaev and A.~A.~Tseytlin,
Nucl. Phys. {\bf B293} (1987) 385.}

\lref\calkp {C.~G.~Callan,
I.~R.~Klebanov and M.~J.~Perry, Nucl. Phys. {\bf B278} (1986)
78.}

\lref\lov {C.~Lovelace, Phys. Lett. {\bf B135} (1984) 75.}

\lref\friv {B.~E.~Fridling and A.~E.~M.~Van de Ven,
Nucl. Phys. {\bf B268} (1986) 719.}

\lref\gepw {D.~Gepner and E.~Witten, Nucl. Phys. {\bf B278}
(1986) 493.}

\lref\din {M.~Dine, Lectures delivered at
TASI 1988, Brown University (1988) 653.}

\lref\berdone {E.~A.~Bergshoeff and M.~de Roo, Phys. Lett.
{\bf B218} (1989)
210.}

\lref\calhs{C.~G.~Callan, J.~A.~Harvey and A.~Strominger,
Nucl. Phys.
{\bf B359} (1991) 611.}

\lref\calhsone{C.~G.~Callan, J.~A.~Harvey and A.~Strominger,
Nucl. Phys.
{\bf B367} (1991) 60.}

\lref\thoone{G.~'t~Hooft, Phys. Rev. Lett. {\bf 37} (1976) 8.}

\lref\wil{F.~Wilczek, in
{\it Quark confinement and field theory},
Eds. D.~Stump and D.~Weingarten, (John Wiley and Sons, New York,
1977).}

\lref\jacnr{R.~Jackiw, C.~Nohl and C.~Rebbi, Phys. Rev.
{\bf D15} (1977)
1642.}

\lref\khuinst{R.~R.~Khuri, Phys. Lett.
{\bf B259} (1991) 261.}

\lref\khumant{R.~R.~Khuri, Nucl. Phys.
 {\bf B376} (1992) 350.}

\lref\khumono{R.~R.~Khuri,
 Phys. Lett. {\bf B294} (1992) 325.}

\lref\khumonscat{R.~R.~Khuri,
Phys. Lett. {\bf B294} (1992) 331.}

\lref\khumonex{R.~R.~Khuri,
Nucl. Phys. {\bf B387} (1992) 315.}

\lref\khumonin{R.~R.~Khuri,
 Phys. Rev. {\bf D46} (1992) 4526.}

\lref\khugeo{R.~R.~Khuri,
Phys. Lett. {\bf 307} (1993) 302.}

\lref\khuscat{R.~R.~Khuri,
 Nucl. Phys. {\bf B403} (1993) 335.}

\lref\khuwind {R.~R.~Khuri, Phys. Rev. {\bf D48} (1993) 2823.}

\lref\gin{P.~Ginsparg, Lectures delivered at
Les Houches summer session, June 28--August 5, 1988.}

\lref\alljj{R. W. Allen, I. Jack and D. R. T. Jones,
 Z. Phys. {\bf C41}
(1988) 323.}

\lref\sev{A. Sevrin, W. Troost and A. van Proeyen,
Phys. Lett. {\bf B208} (1988) 447.}

\lref\schout{K. Schoutens, Nucl. Phys. {\bf B295} [FS21] (1988)
634.}

\lref\harl{J.~A.~Harvey and J.~Liu, Phys. Lett. {\bf B268}
(1991) 40.}

\lref\man{N.~S.~Manton, Nucl. Phys. {\bf B126} (1977) 525.}

\lref\manone{N.~S.~Manton, Phys. Lett. {\bf B110} (1982) 54.}

\lref\mantwo{N.~S.~Manton, Phys. Lett. {\bf B154} (1985) 397.}

\lref\atihone{M.~F.~Atiyah and N.~J.~Hitchin, Phys. Lett.
{\bf A107}
(1985) 21.}

\lref\atihtwo{M.~F.~Atiyah and N.~J.~Hitchin, {\it The Geometry
and
Dynamics of Magnetic Monopoles}, (Princeton University Press,
1988).}

\lref\polc{J.~Polchinski, Phys. Lett. {\bf B209} (1988) 252.}

\lref\gibhp{G.~W.~Gibbons and S.~W.~Hawking, Phys. Rev.
{\bf D15}
(1977) 2752.}

\lref\gibhpone{G.~W.~Gibbons, S.~W.~Hawking and M.~J.~Perry,
 Nucl. Phys.
{\bf B318} (1978) 141.}

\lref\brih{D.~Brill and G.~T.~Horowitz, Phys. Lett. {\bf B262}
(1991)
437.}

\lref\gids{S.~B.~Giddings and A.~Strominger, Nucl. Phys.
{\bf B306}
(1988) 890.}

\lref\gidsone{S.~B.~Giddings and A.~Strominger, Phys. Lett.
{\bf B230}
(1989) 46.}

\lref\canhsw{P.~Candelas, G.~T.~Horowitz, A.~Strominger and
E.~Witten,
Nucl. Phys. {\bf B258} (1984) 46.}

\lref\bog{E.~B.~Bogomolnyi, Sov. J. Nucl. Phys. {\bf 24} (1976)
449.}

\lref\war{R.~S.~Ward, Comm. Math. Phys. {\bf 79} (1981) 317.}

\lref\warone{R.~S.~Ward, Comm. Math. Phys. {\bf 80} (1981) 563.}

\lref\wartwo{R.~S.~Ward, Phys. Lett. {\bf B158} (1985) 424.}

\lref\grop{D.~J.~Gross and M.~J.~Perry, Nucl. Phys. {\bf B226}
(1983)
29.}

\lref\ash{{\it New Perspectives in Canonical Gravity}, ed.
A.~Ashtekar,
(Bibliopolis, 1988).}

\lref\lic{A.~Lichnerowicz, {\it Th\' eories Relativistes de la
Gravitation et de l'Electro-magnetisme}, (Masson, Paris 1955).}

\lref\gol{H.~Goldstein, {\it Classical Mechanics},
Addison-Wesley,
1981.}

\lref\ros{P.~Rossi, Physics Reports, 86(6) 317-362.}

\lref\dixdp{J.~A.~Dixon, M.~J.~Duff and J.~C.~Plefka,
 Phys. Rev.
Lett.
{\bf 69} (1992) 3009.}

\lref\chad{J.~M.~Charap and M.~J.~Duff, Phys. Lett. {\bf B69}
(1977) 445.}

\lref\dufkexst{M.~J.~Duff and R.~R.~Khuri,
Nucl. Phys. {\bf B411} (1994) 473.}

\lref\khubifb{R.~R.~Khuri,
Phys. Rev. {\bf D48} (1993) 2947.}

\lref\khustab{R.~R.~Khuri, Phys. Lett. {\bf B307} (1993) 298.}

\lref\sor{R.~D.~Sorkin, Phys. Rev. Lett. {\bf 51} (1983) 87.}

\lref\dabh{A.~Dabholkar and J.~A.~Harvey,
 Phys. Rev. Lett. {\bf 63} (1989) 478.}

\lref\fels{A.~G.~Felce and T.~M.~Samols, Phys. Lett.
{\bf B308} (1993) 30.}

\lref\dufipss{M.~J.~Duff, T.~Inami, C.~N.~Pope, E.~Sezgin and
K.~S.~Stelle,
Nucl. Phys. {\bf B297}
(1988) 515.}

\lref\fujku{K.~Fujikawa and J.~Kubo,
 Nucl. Phys. {\bf B356} (1991) 208.}

\lref\cvet{M.~Cveti\v c, Phys. Rev. Lett. {\bf 71} (1993) 815.}

\lref\cvegs{M.~Cveti\v c, S. Griffies and H. H. Soleng, Phys.
 Rev. Lett. {\bf 71} (1993) 670; Phys. Rev. {\bf D48} (1993)
 2613.}

\lref\la{H. S. La, Phys. Lett. {\bf B315} (1993) 51.}

\lref\gresvy{B.~R.~Greene, A.~Shapere, C.~Vafa and S.~T.~Yau,
 Nucl. Phys.
{\bf B337} (1990) 1.}

\lref\fonilq{A.~Font, L.~Ib\'a\~nez, D.~Lust and F.~Quevedo,
Phys. Lett.
{\bf B249} (1990) 35.}

\lref\bin{P.~Bin\'etruy, Phys. Lett. {\bf B315} (1993) 80.}

\lref\koun{C.~Kounnas, in {\it Proceedings of INFN Eloisatron
Project, 26th Workshop: ``From Superstrings to Supergravity",
 Erice, Italy,
Dec. 5-12, 1992}, Eds. M.~Duff, S.~Ferrara and R.~Khuri,
(World Scientific, 1994).}

\lref\duftv{M.J. Duff, P.K. Townsend and P. van Nieuwenhuizen,
Phys. Lett. {\bf B122} (1983) 232.}

\lref\dufgt{M.J. Duff, G.W. Gibbons and P.K. Townsend
Phys. Lett. {\bf B} (1994) }

\lref\guv{R. G\"uven, Phys. Lett. {\bf B276} (1992) 49.}

\lref\guven{R. G\"uven, Phys. Lett. {\bf B212} (1988) 277.}

\lref\dobm{P.~Dobiasch and D.~Maison, Gen. Rel. Grav.
{\bf 14} (1982) 231.}

\lref\chod{A.~Chodos and S.~Detweiler, Gen. Rel. Grav.
{\bf 14} (1982) 879.}

\lref\pol{D.~Pollard, J. Phys. {\bf A16} (1983) 565.}

\lref\duffkr{M.~J.~Duff, S.~Ferrara, R.~R.~Khuri and J.~Rahmfeld,
 in preparation.}

\lref\lu{J. X. Lu,
 Phys. Lett. {\bf B313} (1993) 29.}

\lref\grel{R. Gregory and R. Laflamme, Phys. Rev. Lett.
{\bf 70} (1993) 2837.}

\lref\rom{L. Romans, Nucl. Phys. {\bf B276} (1986) 71.}

\lref\sala {A. Salam and E. Sezgin, {\it Supergravities in
Diverse Dimensions},
(North Holland/World Scientific, 1989).}

\lref\strath {J. Strathdee, Int. J. Mod. Phys. {\bf A2}
(1987) 273.}

\lref\dufliib{M.~J.~Duff and J.~X.~Lu,
 Nucl. Phys. {\bf B390} (1993) 276.}

\lref\nictv{H. Nicolai, P. K. Townsend and P. van
Nieuwenhuizen, Lett. Nuovo
Cimento {\bf 30} (1981) 315.}

\lref\towspan{P. K. Townsend, in Proceedings of the 13th GIFT
Seminar
on Theoretical Physics: {\it Recent Problems in Mathematical
Physics} Salamanca, Spain, 15-27 June, 1992.}

\lref\dufm{M.~J.~Duff and R.~Minasian,
 Nucl. Phys. {\bf B436} (1995) 507.}

\lref\gropy{D. J. Gross, R. D. Pisarski and L. G. Yaffe,
Rev. Mod. Phys.
{\bf 53} (1981) 43.}

\lref\rohw{R. Rohm and E. Witten,
 Ann. Phys. {\bf 170} (1986) 454.}

\lref\banddf{T. Banks, M. Dine, H. Dijkstra and W. Fischler,
Phys. Lett. {\bf B212} (1988) 45.}

\lref\ferkp{S. Ferrara, C. Kounnas and M. Porrati, Phys. Lett.
{\bf B181} (1986) 263.}

\lref\ter{M. Terentev, Sov. J. Nucl. Phys. {\bf 49} (1989) 713.}

\lref\hass{S. F. Hassan and A. Sen, Nucl. Phys. {\bf B375}
(1992) 103.}

\lref\mahs{J. Maharana and J. Schwarz, Nucl. Phys. {\bf B390}
(1993) 3.}

\lref\senrev{A. Sen,
 Int. J. Mod. Phys. {\bf A9} (1994) 3707.}

\lref\senone{A.~Sen,
Nucl. Phys. {\bf B404} (1993) 109.}

\lref\sentwo{A.~Sen,
  Int. J. Mod. Phys. {\bf A8} (1993) 5079.}

\lref\schsen{J.~H.~Schwarz and A.~Sen,
Nucl. Phys. {\bf B411} (1994) 35.}

\lref\schsentwo{J.~H.~Schwarz and A.~Sen,
 Phys. Lett. {\bf B312} (1993) 105.}

\lref\schtwo{J.~Schwarz,
CALT-68-1815.}

\lref\senph{A.~Sen, Phys. Lett. {\bf B303} (1993) 22.}

\lref\schwarz{J.~H.~Schwarz, CALT-68-1879, hepth/9307121.}

\lref\dufldr{M. J. Duff and J. X. Lu, Nucl. Phys. {\bf B347}
(1990) 394.}

\lref\salstr{A. Salam and J. Strathdee, Phys. Lett. {\bf B61}
(1976) 375.}

\lref\jjj{J. Gauntlett, J. Harvey and J. T. Liu,
Nucl. Phys. {\bf B409} (1993) 363.}

\lref\dupo{M.~J.~Duff and C.~N.~Pope, Nucl. Phys {\bf B255}
(1985) 355.}

\lref\sg{E.~Cremmer, S.~Ferrara, L.~Girardello and
 A.~Van Proeyen,
 Phys. Lett. {\bf B116} (1982) 231.}

\lref\sgone{E.~Cremmer, S.~Ferrara, L.~Girardello and
 A.~Van Proeyen,
 Nucl. Phys. {\bf B212} (1983) 413.}

\lref\fkp{S. Ferrara, C. Kounnas and M. Porrati,
Phys. Lett. {\bf B181} (1986) 263.}

\lref\fp{S. Ferrara and M. Porrati,
Phys. Lett. {\bf B216} (1989) 289.}

\lref\ghs{D.~Garfinkle, G.~T.~Horowitz and A.~Strominger,
Phys. Rev. {\bf D43}
(1991) 3140.}

\lref\hor{G.~T.~Horowitz, in Proceedings of Trieste '92,
{\it String theory and quantum gravity '92} p.55.}

\lref\gidps{S.~B.~Giddings, J.~Polchinski and A.~Strominger,
Phys. Rev. {\bf D48} (1993) 5784.}

\lref\shatw{A.~Shapere, S.~Trivedi and F.~Wilczek,
 Mod. Phys. Lett. {\bf A6}
(1991) 2677.}

\lref\klopv{R.~Kallosh, A.~Linde, T.~Ortin, A.~Peet and
A.~Van~Proeyen,
       Phys. Rev. {\bf D46} (1992) 5278.}

\lref\kal{R.~Kallosh, Phys. Lett. {\bf B282} (1992) 80.}

\lref\ko{R.~Kallosh and T.~Ortin, Phys. Rev. {\bf D48} (1993)
742.}

\lref\hw{C.~F.~E.~Holzhey and F.~Wilczek, Nucl. Phys.
{\bf B360} (1992) 447.}

\lref\dauria{R.~D'Auria, S.~Ferrara and M.~Villasante,
Class. Quant. Grav. {\bf 11} (1994) 481.}

\lref\gibp{G.~W.~Gibbons and M.~J.~Perry, Nucl. Phys.
{\bf B248} (1984) 629.}

\lref\salam{S. W. Hawking, Monthly Notices Roy. Astron. Soc.
 {\bf 152} (1971) 75; Abdus Salam in
   {\it Quantum Gravity: an Oxford Symposium} (Eds. Isham,
Penrose
and Sciama, O.U.P. 1975); G. 't Hooft, Nucl. Phys. {\bf B335}
(1990) 138.}

\lref\susskind{ L. Susskind, RU-93-44, hepth/9309145;
   J. G. Russo and L. Susskind, UTTG-9-94,
hepth/9405117.}

\lref\gibbons{ G. W. Gibbons, in {\it Supersymmetry,
Supergravity
and Related Topics}, Eds. F. del Aguila, J. A. Azcarraga and
L. E. Ibanez
(World Scientific, 1985).}

\lref\aichelburg{ P. Aichelburg and F. Embacher, Phys. Rev.
{\bf D37}
(1986) 3006.}

\lref\geroch{ R. Geroch, J. Math. Phys. {\bf 13} (1972) 394.}

\lref\hosoya{ A. Hosoya, K.
Ishikawa, Y. Ohkuwa and K. Yamagishi, Phys. Lett. {\bf B134}
(1984) 44.}

\lref\gibw{ G. W. Gibbons
and D. L. Wiltshire, Ann. of Phys. {\bf 167} (1986) 201.}

\lref\senprl{ A. Sen, Phys. Rev. Lett. {\bf 69}
(1992) 1006.}

\lref\schild{ G. C. Debney, R. P. Kerr and
   A. Schild, J. Math. Phys. {\bf 10} (1969) 1842.}

\lref\hort{G. T. Horowitz and A. A. Tseytlin, Phys. Rev.
{\bf D50} (1994) 5204.}

\lref\hortsey{G. T. Horowitz and A. A. Tseytlin,
Imperial/TP/93-94/51, UCSBTH-94-24, hepth/9408040;
Imperial/TP/93-94/54, UCSBTH-94-31, hepth/9409021.}

\lref\cvey{M. Cveti\v c and D. Youm, UPR-623-T, hepth/9409119.}

\lref\tseytlin{A. A. Tseytlin, Imperial-TP-93-94-46,
hepth/9407099.}

\lref\klim{C. Klimcik and A. A. Tseytlin, Nucl. Phys.
{\bf B424} (1994) 71.}

\lref\witdiv{E. Witten, IASSNS-HEP-95-18, hepth/9503124.}

\lref\harvstrom{J. Harvey and A. Strominger, EFI-95-16,
  hepth/9504047.}

\lref\sensix{A. Sen, TIFR-TH-95-16, hepth/9504027.}

\lref\vafwit{C. Vafa and E. Witten, HUTP-95-A015, hepth/9505053.}

\lref\hulttwo{C. M. Hull and P. K. Townsend, QMW-95-10,
hepth/9505073.}

\lref\drtwo{M. J. Duff and J. Rahmfeld, in preparation.}

\lref\duduals{M. J. Duff, Nucl. Phys {\bf B442} (1995) 47.}

\lref\sengeroch{A. Sen, TIFR-TH-95-10, hepth/9503057.}


\newsec{Introduction}

The existence of large classes of soliton solutions in string
theory is intimately connected with the presence of various
dualities in string theory (for recent reviews of string
solitons, see \refs{\prep,\bhs}).
Most of the solitonic solutions found
so far break half of the spacetime supersymmetries
of the theory in which they arise. Examples
of string-like solitons
in this class are
the fundamental string solution of \dabghr\ and the dual
string solution of \dufkexst, which are interchanged once
the roles of the strong/weak coupling
$S$-duality and target space $T$-duality are interchanged.
For $N=4$, $D=4$ compactifications of heterotic string theory,
$T$-duality corresponds to the discrete group $O(6,22;Z)$
and is known to be an
exact symmetry of the full string theory. For a review, see
\refs{\givpr}.
There is now a good deal of evidence in favour of the
conjecture
that
the $S$-duality group $SL(2,Z)$ is also an exact symmetry of
the full string
theory. For a review, see
\refs{\senrev}.
The dual string of \dufkexst\
thus belongs to an $O(6,22;Z)$ family of dual
strings just as
there is an $SL(2,Z)$ family of fundamental strings \sentwo.
This accords with
the observation  that string/string duality
interchanges the roles of strong/weak coupling duality and
target space
duality \duduals. For earlier
discussions of the string/string duality
conjecture see
\refs{\duflloop,\dufkexst,\duflblacks\dufm{--}\glasgow,\prep};
for recent ramifications see
\refs{\hulttwo\witdiv\harvstrom\sensix{--}\vafwit}.

Interesting examples of solutions which break more than
half of the spacetime supersymmetries are
the $D=10$ double-instanton string soliton of \khubifb\ (which
breaks
3/4), the $D=10$
octonionic string soliton of \hars\ (which breaks
15/16) and the $D=11$ extreme black fourbrane and sixbrane
of \guv\ (which break 3/4 and 7/8 respectively).
In this paper
we present new classes of string-like solutions which
arise in heterotic string theory toroidally compactified
to four dimensions. Connections are made between the
solution-generating subgroup of the
$T$-duality group and the number of spacetime
supersymmetries broken in the $N=4$ theory. Analogous
solutions are also seen to arise in $N=2$ and $N=1$
compactifications. Recent discussions of
supersymmetry and duality can be found
in \refs{\bakone\baktwo\bakthree\cvey\bko{--}\baksfet}.

Next, the conjecture \refs{\dufldr\dufr\sentd{--}\hult}
that $S$- and $T$-duality can be
united into
$O(8,24;Z)$ is discussed. In a recent paper by Sen \sentd, the
fundamental string is related to the stringy cosmic string
\gresvy\ by an $O(8,24;Z)$ transformation. In this paper,
we find
an $O(8,24;Z)$ transformation relating the fundamental string
to the dual string of \dufkexst, thus supporting the
above conjecture.

Finally, we speculate on the significance of these solutions to
string/string duality.

\newsec{Generalized $T$ Solutions in $N=4$}

We adopt the following conventions for
$N=1$, $D=10$ heterotic string theory compactified to
$N=4$, $D=4$
heterotic string theory: $(0123)$ is the
four-dimensional spacetime, $z=x_2+ix_3=re^{i\theta}$,
$(456789)$ are
the compactified directions, $S=e^{-2\Phi} + ia=S_1+iS_2$,
where
$\Phi$ and $a$ are the four-dimensional dilaton and axion and
\eqn\tdefn{\eqalign{T^{(1)}&=T^{(1)}_1+iT^{(1)}_2=
\sqrt{{\rm det}
g_{mn}}-iB_{45}, \quad\quad m,n=4,5,\cr
T^{(2)}&=T^{(2)}_1+iT^{(2)}_2=\sqrt{{\rm det} g_{pq}}-iB_{67},
\quad\quad p,q=6,7,\cr
T^{(3)}&=T^{(3)}_1+iT^{(3)}_2=\sqrt{{\rm det} g_{rs}}-iB_{89},
\quad\quad r,s=8,9\cr}}
are the moduli. Throughout this section,
and unless specified otherwise in the rest of the paper,
 we assume dependence only on the
coordinates $x_2$ and $x_3$ (i.e. $x_0$ and $x_1$ are
Killing directions), and that no other moduli than the ones
above
are nontrivial.

The canonical four-dimensional bosonic action for the
above compactification ansatz in the
gravitational sector can be written in terms of $g_{\mu\nu}$
($\mu,\nu=0,1,2,3$), $S$ and $T^{(a)}, a=1,2,3$ as
\eqn\sfour{\eqalign{S_4=\int d^4x \sqrt{-g}\biggl(&R-
{g^{\mu\nu}\over
2S^2_1}\partial_\mu S \partial_\nu \bar S \cr &-
{g^{\mu\nu}\over
2T^{(1)^2}_1}\partial_\mu T^{(1)}
\partial_\nu \bar T^{(1)} - {g^{\mu\nu}\over
2T^{(2)^2}_1}\partial_\mu T^{(2)}
\partial_\nu \bar T^{(2)} - {g^{\mu\nu}\over
2T^{(3)^2}_1}\partial_\mu T^{(3)}
\partial_\nu \bar T^{(3)} \biggr).\cr}}
A solution for this action for $S=1$ ($\Phi=a=0$) is given
by
the metric
\eqn\metsol{ds^2=-dt^2+dx_1^2 + T^{(1)}_1 T^{(2)}_1 T^{(3)}_1
(dx_2^2+dx_3^2),}
where three cases with different nontrivial $T$-duality arise
 depending on
the
number $n$ of nontrivial $T$ moduli:
\eqn\tsol{\eqalign{n&=1: \qquad\qquad T^{(1)}=-{1\over 2\pi}\ln
{z\over r_0},
\qquad T^{(2)}=T^{(3)}=1,\cr
n&=2: \qquad\qquad T^{(1)}=T^{(2)}=-{1\over 2\pi}\ln
{z\over r_0},
\qquad T^{(3)}=1,\cr
n&=3: \qquad\qquad T^{(1)}=T^{(2)}=T^{(3)}=-{1\over 2\pi}
\ln {z\over
r_0}.\cr}}
In each of the expressions for $T^{(a)}$, $z$ may be replaced by
$\bar
z$ independently (i.e. there is a freedom in changing
 the sign of the
axionic part of the modulus).
Note that the $n=1$ case is simply the dual
string solution of \dufkexst.
Since $S_4$ has manifest $SL(2,R)$ duality in each of the moduli
(broken to $SL(2,Z)$ in string theory), we
can generate from the $n=2$ case an $SL(2,Z)^2$ family of
 solutions
and from the $n=3$ case an $SL(2,Z)^3$ family of solutions.
Note that the full $T$-duality group in all three cases
remains $O(6,22;Z)$, but that the subgroup with nontrivial
action on the particular solutions (or the solution-generating
subgroup referred to above) for $n=1,2,3$ is given
by $SL(2,Z)^n$ \refs{\fkp,\fp}.

{}From the ten-dimensional viewpoint, the $n=3$ solution, for
example, can be rewritten in the string sigma-model metric
 frame as
\eqn\tend{\eqalign{e^{2\phi}&=(-{1\over 2\pi}
\ln {r\over r_0})^3,\cr
ds^2&=-dt^2+dx_1^2 + e^{2\phi} (dx_2^2+dx_3^2) +
e^{2\phi/3} (dx_4^2 + ...+ dx_9^2),\cr
B_{45}&=\pm B_{67}=\pm B_{89}=\pm {\theta\over 2\pi},\cr}}
where $\phi$ is the ten-dimensional
dilaton.

The solution \tsol\ can in fact be generalized to include an
arbitrary number of string-like sources in each $T^{(i)}$
\eqn\threet{\eqalign{
ds^2&=-dt^2+dx_1^2 + T^{(1)}_1 T^{(2)}_1 T^{(3)}_1
(dx_2^2+dx_3^2) \cr
 T^{(1)}& =
-{1\over 2\pi} \sum_{j=1}^M m_j\ln {(z-b_j)\over r_{j0}} ,\cr
T^{(2)}& =
-{1\over 2\pi} \sum_{k=1}^P p_k\ln {(z-c_k)\over r_{k0}} ,\cr
T^{(3)}& =
-{1\over 2\pi} \sum_{l=1}^Q q_l\ln {(z-d_l)\over r_{l0}} ,\cr}}
where  $M, P$ and $Q$ are
arbitrary numbers of string-like solitons in
$T^{(1)}, T^{(2)}$ and $T^{(3)}$ respectively
each with arbitrary location $b_j, c_k$ and $d_l$ locations in
the
complex $z$-plane and arbitrary
winding number $m_j, p_k$ and $q_l$
respectively. The solutions with
 $1,2$ and $3$ nontrivial $T$ fields break $1/2, 3/4$ and $7/8$
of the spacetime supersymmetries respectively. Again, one can
make the replacement $z\to \bar z$ independently
in each of the moduli, so that each $T^{(i)}$ is either
analytic or anti-analytic in $z$.

\newsec{Supersymmetry Breaking}

We claim that the above solutions for the massless fields in
the
gravitational sector when combined with a Yang-Mills field
 given by
$A_M^{PQ}=\Omega_M^{PQ}=\omega_M^{PQ} \pm 1/2 H_M{}^{PQ}$ (the
usual expedient of equating the
gauge to the generalized connection) solve the tree-level
supersymmetry equations of the heterotic string for zero fermi
fields
and can be argued to be exact solutions of heterotic string
theory \refs{\calhs,\dufkexst}.
The supersymmetry equations in $D=10$ are given by
\eqn\sseq{\eqalign{\delta\psi_M&=\left(\partial_M+{\textstyle
{1\over
4}}\Omega_{MAB}
\Gamma^{AB}\right)\epsilon=0,\cr
\delta\lambda&=\left(\Gamma^A\partial_A\phi-{\textstyle{1\over
12}}
H_{ABC}\Gamma^{ABC}\right)\epsilon=0,\cr
\delta\chi&=F_{AB}\Gamma^{AB}\epsilon=0, \cr}}
where $A,B,C,M=0,1,2,...,9$ and
where $\psi_M,\ \lambda$ and $\chi$ are the gravitino, dilatino
and
gaugino
fields. The Bianchi identity is given by
\eqn\bianchi{dH={\alpha'\over 4} \left({\rm tr} R\wedge R-
{1\over
30}{\rm Tr}
F\wedge F\right),}
and is satisfied automatically for this ansatz. We further claim
that
the $n=1,2,3$ solutions break $1/2, 3/4$ and $7/8$ of the
spacetime
supersymmetries respectively. We will show this to be true for
the most general case of $n=3$.

$\delta\lambda=0$ follows from scaling, since the dilaton can
be
written as the sum of three parts (the moduli) each of which
produces
a contribution which cancels against the contribution of the
$H$ term
coming from the appropriate
four-dimensional subspace. In other words, each of the
subspaces
$(2345)$, $(2367)$ and $(2389)$ effectively contains a
four-dimensional axionic  (anti) instanton
\refs{\reyone,\khuinst,\khumonin}\ with the appropriate (anti)
self-duality in the
generalized connection in the respective
subspace, depending on whether the corresponding
modulus is analytic or anti-analytic in $z$.
 Another way of saying this is that there are three
independent parts of $\delta\lambda$, each of which vanishes
as in the simple $n=1$
case, for the appropriate chirality choice of $\epsilon$ in the
respective four-dimensional subspace.

$\delta\psi_M=0$ is a little more subtle. For the $n=1$
case,
the generalized connection is an
instanton \refs{\khuinst,\khumonin}, and for constant
chiral spinor $\epsilon$ with
chirality
in the
four-space of the instanton opposite to that of the instanton
(e.g.
negative for instanton and positive for anti-instanton), it is
easy
to show that
$\Omega_M^{AB}\Gamma_{AB}\epsilon=0$. In the more general
$n=3$ case, we
proceed as follows. It is sufficient to show that
$\delta\psi_M=0$
for $M=2$ and $M=4$
(i.e. for a spacetime and for a compactified index),
as for $M=0,1$ the supersymmetry variation is trivial,
while for the rest of the indices the arguments are identical
to one of the above two representative cases. For $M=2$
this
can be written out explicitly as
\eqn\psitwo{\eqalign{4\delta\psi_2=&\left({1\over 3}
\omega_2^{23}
\Gamma_{23} +
\omega_2^{24} \Gamma_{24} + \omega_2^{25} \Gamma_{25} -
{1\over 2}
H_2{}^{45}
\Gamma_{45}\right) \epsilon \cr
+  &\left({1\over 3} \omega_2^{23} \Gamma_{23} +
\omega_2^{26} \Gamma_{26} + \omega_2^{27} \Gamma_{27} -
{1\over 2}
H_2{}^{67}
\Gamma_{67}\right) \epsilon \cr
+  &\left({1\over 3} \omega_2^{23} \Gamma_{23} +
\omega_2^{28} \Gamma_{28} + \omega_2^{29} \Gamma_{29} -
{1\over 2}
H_2{}^{89}
\Gamma_{89}\right) \epsilon .\cr}}
Each line in \psitwo\ acts on only a four-dimensional component
of $\epsilon$ and can be shown to exactly correspond to the
contribution of the supersymmetry equation of a single $n=1$
axionic instanton.
So in effect the configuration carries three such instantons
in the
generalized curvature in the spaces $(2345)$, $(2367)$ and
$(2389)$.
Therefore for the appropriate chirality of the four-dimensional
components of $\epsilon$ (depending on the choices of
analyticity
of the $T$ fields),
$\delta\psi_2=0$. Since we are making three such choices, $1/8$
of the spacetime supersymmetries are preserved and $7/8$
are broken. Another, perhaps
simpler, way to understand this is to write
$\epsilon=\epsilon_{(01)}\otimes \epsilon_{(23)}\otimes
\epsilon_{(45)}\otimes
\epsilon_{(67)}\otimes \epsilon_{(89)}$. Then the chiralities of
$\epsilon_{(45)}, \epsilon_{(67)}$ and $\epsilon_{(89)}$ are all
correlated with that
of $\epsilon_{(23)}$, so it follows that $7/8$ of the
supersymmetries
are broken.

We also need to check $\delta\psi_4=0$. In this case, it is
easy to
show that the whole term reduces exactly to the contribution of
a single $n=1$ axionic instanton:
\eqn\psifour{4\delta\psi_4= \left(\omega_4^{42} \Gamma_{42} +
\omega_4^{43} \Gamma_{43} - {1\over 2} H_4{}^{25}
\Gamma_{25} - {1\over 2} H_4{}^{35}
\Gamma_{35}\right)\epsilon =0,}
as in this case there is only the contribution of the instanton
in the $(2345)$ subspace. $\delta\psi_4$ then vanishes for the
same chirality choice of $\epsilon$ as in the paragraph
above.

There remains to show that $\delta\chi=0$. This can be easily
seen by
noting that, as in the $\delta\psi_M$ case, the term
$F_{23}\Gamma^{23}$ splits into three equal pieces, each of
which
combines with the rest of a $D=4$ instanton
(since the Yang-Mills connection is equated to the generalized
connection and is also effectively an instanton in each of the
three
four-dimensional subspaces) to give a zero contribution for
the same chirality choices in the four-dimensional
subspaces as above.

For the $n=2$ case, it is even easier to show that $3/4$ of the
supersymmetries are broken. Tree-level neutral versions
($A_M=0$) of these
solutions also follow immediately and reduce to \metsol\
and \tsol\ on compactification to $D=4$, where, of course,
the same degree of supersymmetry breaking for each
class of solutions may be verified
directly. Henceforth we will consider only neutral solutions.

\newsec{Generalized Solutions with Nontrivial $S$}

It turns out that these solutions generalize even further to
solutions which include a nontrivial $S$ field. The net result
of
adding a nontrivial $S$ (with $SL(2,Z)$ symmetry) is to break
half again of the remaining spacetime supersymmetries
preserved by the corresponding $T$ configuration with trivial
$S$, except for the case of $n=3$ nontrivial moduli, which is a
bit
more subtle and will be discussed below.
In particular, the simplest solution of the action
\sfour\ with one nontrivial $S$ and three nontrivial $T$
moduli has the form
\eqn\sthreet{\eqalign{ds^2&=-dt^2+dx_1^2
+ S_1  T_1^{(1)}  T_1^{(2)}  T_1^{(3)} (dx_2^2 + dx_3^2),\cr
S&=T^{(1)}=T^{(2)}=T^{(3)}=-{1\over 2\pi} \ln {z\over r_0},\cr}}
where again we have an $SL(2,Z)$ symmetry in $S$ and
in each of the
$T$ fields.

It is interesting to note that the real parts of the $S$ and
$T$ fields can
be arbitrary as long as they satisfy the box equation in
the two-dimensional subspace $(23)$. In particular, each can be
generalized to multi-string configurations independently,
with arbitrary number of strings each with
arbitrary winding number. The
corresponding imaginary part can most easily be found by going
to ten dimensions, where the corresponding $B$-field follows
from the modulus. So there is nothing special about the choice
$\ln z$. It is merely the simplest case.

The ten-dimensional form
of the most general solution can be written in the string
sigma-model metric frame as
\eqn\sthreetten{\eqalign{
ds^2&=e^{2\Phi}(-dt^2+dx_1^2) + e^{2(\sigma_1 + \sigma_2 +
\sigma_3)} (dx_2^2 + dx_3^2) \cr
 &+ e^{2\sigma_1} (dx_4^2 + dx_5^2)
+  e^{2\sigma_2} (dx_6^2 + dx_7^2) +
 e^{2\sigma_3} (dx_8^2 + dx_9^2), \cr
S&=e^{-2\Phi} + ia = -{1\over 2\pi} \sum_{i=1}^N n_i
\ln {(z-a_i)\over r_{i0}} ,\cr
T^{(1)}&=e^{2\sigma_1} - iB_{45} =
-{1\over 2\pi} \sum_{j=1}^M m_j\ln {(z-b_j)\over r_{j0}} ,\cr
T^{(2)}&=e^{2\sigma_2} - iB_{67} =
-{1\over 2\pi} \sum_{k=1}^P p_k\ln {(z-c_k)\over r_{k0}} ,\cr
T^{(3)}&=e^{2\sigma_3} - iB_{89} =
-{1\over 2\pi} \sum_{l=1}^Q q_l\ln {(z-d_l)\over r_{l0}} ,\cr
\phi&=\Phi +\sigma_1 + \sigma_2 + \sigma_3,\cr}}
where $\phi$ is the ten-dimensional dilaton, $\Phi$ is the
four-dimensional dilaton, $\sigma_i$ are the metric
moduli, $a$ is the axion in the
four-dimensional subspace $(0123)$ and $N, M, P$ and $Q$ are
arbitrary numbers of string-like solitons in
$S, T^{(1)}, T^{(2)}$ and $T^{(3)}$ respectively
each with arbitrary location $a_i, b_j, c_k$ and $d_l$ in the
complex
$z$-plane and
arbitrary winding number $n_i, m_j, p_k$ and $q_l$
respectively. Again one can replace $z$ by $\bar z$ independently
in $S$ and in each of the moduli.

The solutions with nontrivial $S$ and $0, 1$ and $2$ nontrivial
$T$ fields preserve $1/2, 1/4$  and $1/8$ spacetime
supersymmetries
respectively.
This follows from the fact that the nontrivial $S$ field breaks
half of the remaining supersymmetries by imposing a chirality
choice on the spinor $\epsilon$ in the
$(01)$ subspace of the ten-dimensional space.
The solution with nontrivial $S$
and $3$ nontrivial $T$ fields breaks $7/8$ of the spacetime
supersymmetries for one chirality choice of $S$, and all the
spacetime supersymmetries for the other. This can be seen as
follows:
the three nontrivial $T$ fields, when combined with an overall
chirality choice of the Majorana-Weyl spinor in ten dimensions,
impose a chirality choice on $\epsilon_{01}$. If this choice
agrees with the chirality choice imposed by $S$, then no
more supersymmetries are broken, and so $1/8$ are preserved
(or $7/8$ are broken). When these two choices are not identical,
all the supersymmetries are broken, although the ansatz remains
a solution to the bosonic action.

\newsec{Dyonic Strings in $D=6$ and $D=10$}

A special case of the above generalized $S$ and $T$
 solutions is the one
with nontrivial $S$ and only one nontrivial $T$.
This is in fact a ``dyonic'' solution which interpolates
between the fundamental $S$ string of \dabghr\ and the
dual $T$ string of \dufkexst. It turns out that in going to
higher
dimensions, one still has a solution even if the box
equation covers the whole transverse four-space $(2345)$
(the remaining four directions are flat even in $D=10$, as
$\sigma_2=\sigma_3=0$).
The $D=10$ form in fact reduces to a $D=6$ dyonic solution
($i=2,3,4,5$)
\eqn\dyon{\eqalign{\phi&=\Phi_E + \Phi_M, \cr
ds^2&=e^{2\Phi_E} (-dt^2+dx_1^2)
+ e^{2\Phi_M} dx_i dx^i,\cr
e^{-2\Phi_E}&=1+{Q_E\over y^2},\qquad\qquad
e^{2\Phi_M}=1+{Q_M\over y^2},\cr
H_3&=2Q_M\epsilon_3,\qquad\qquad
e^{-2\phi}{}{} ^* H_3=2Q_E\epsilon_3\cr}}
for the special case of a single electric and single
magnetic charge at $y=0$.
Again this solution generalizes to one with an arbitrary
number of arbitrary (up to dyonic
quantization conditions) charges
at arbitrary locations in the transverse four-space.
 For $Q_M=0$ \dyon\ reduces to the solution of
\dabghr\ in $D=6$, while for $Q_E=0$ \dyon\ reduces to the
$D=6$ dual string of \duflblacks\
(which can be obtained from the fivebrane soliton \duflfb\
simply
by compactifying four flat directions). This solution breaks
$3/4$ of the spacetime supersymmetries.
The self-dual limit $Q_E=Q_M$ of this solution has already been
found in \duflblacks\ in the context of $N=2$, $D=6$ self-dual
supergravity, where the solution was shown to
break $1/2$ the spacetime
 supersymmetries. This corresponds precisely to breaking
$3/4$ of the spacetime supersymmetries in the
non self-dual theory in this paper \prep.

Finally, one can generalize the dyonic solution to the
following solution in $D=10$:
\eqn\bibi{\eqalign{
ds^2 &=e^{2\Phi_E}(-dt^2+dx_1^2)
+e^{2\Phi_{M1}}\delta_{ij} dx^i dx^j
+e^{2\Phi_{M2}}\delta_{ab} dx^a dx^b, \cr
\phi &=\Phi_E + \Phi_{M1} + \Phi_{M2},\qquad\qquad
\Phi_E =\Phi_{E1} + \Phi_{E2}, \cr
e^{2\Phi_{E1}} \Box_1 \ e^{-2\Phi_{E1}} &=
e^{2\Phi_{E2}} \Box_2 \ e^{-2\Phi_{E2}}=
e^{-2\Phi_{M1}} \Box_1 \ e^{2\Phi_{M1}}=
e^{-2\Phi_{M2}} \Box_2 \ e^{2\Phi_{M2}}=0, \cr
B_{01} &=e^{2\Phi_E},\qquad
H_{ijk}=2\epsilon_{ijkm}\partial^m \Phi_{M1},\qquad
H_{abc}=2\epsilon_{abcd}\partial^d \Phi_{M1},\cr}}
where $i,j,k,l=2,3,4,5$, $a,b,c,d=6,7,8,9$, $\Box_1$ and
$\Box_2$
represent the Laplacians in the subspaces $(2345)$ and
$(6789)$ respectively and $\phi$ is the ten-dimensional
dilaton. This solution with all fields nontrivial
 breaks $7/8$ of the
spacetime supersymmetries. For $\Phi_{E2}=\Phi_{M2}=0$ we
recover the dyonic solution \dyon\ which breaks $3/4$
of the supersymmetries, for $\Phi_{E1}=\Phi_{E2}=0$
we recover the double-instanton solution of \khubifb\ which
also breaks
$3/4$ of the supersymmetries, while for $\Phi_{M1}=\Phi_{M2}=0$
we obtain the dual of the double-instanton solution, and which,
however, breaks only $1/2$ of the supersymmetries.

\newsec{Solutions in $N=1$ and $N=2$}

It turns out that most of the above solutions that break
$1/2, 3/4, 7/8$ or all of the spacetime supersymmetries in
$N=4$ have analogs in $N=1$ or $N=2$
that break only $1/2$ the spacetime supersymmetries.

For simplicity, let us consider the case of $N=1$,
as the $N=2$ case is similar.
It turns out that the number of nontrivial
$T$ fields with the same analyticity
and the inclusion of a nontrivial $S$ field with the same
analyticity
does
not affect
the number of supersymmetries broken, as in the supersymmetry
equations the contribution of each field is independent. In
particular, the presence of an additional field produces no
new condition on the chiralities, so that the number of
supersymmetries broken is the same for any number of fields,
provided
the fields have the same analyticity or anti-analyticity in $z$,
corresponding to different chirality choices on the
four-dimensional
spinor. This can be seen as follows below.

The supersymmetry transformations in $N=1$ for nonzero
metric, $S$ and moduli fields are given by
\refs{\sg,\sgone}\
\eqn\ssnone{\eqalign{\delta\psi_{\mu L}&=
\left(\partial_\mu + {1\over 2}
\omega_{\mu mn}\sigma^{mn}\right) \epsilon_L
-{\epsilon_L\over 4}\left( {\partial G\over \partial z_i}
\partial_\mu z_i -{\partial  G\over \partial \bar z_i}
\partial_\mu \bar z_i \right)=0, \cr
\delta \chi_{iL}&={1\over 2} \hat D z_i \epsilon_R=0,\cr}}
where $\omega$ is the spin connection,
$\sigma^{mn}=(1/4)[\gamma^m,\gamma^n]$,
$\epsilon_{L,R}=(1/2)(1\pm \gamma^5)\epsilon$,
$\hat D=\gamma^\mu D_{\mu}$, $z_i=S, T^{(1)}, T^{(2)}, T^{(3)}$,
and where
\eqn\gpot{G= -\ln (S+\bar S) - \sum_{j=1}^3 \ln (T^{(j)}+
\bar T^{(j)}).}
Consider the case of a single
nontrivial $T=T^{(1)}$ field (i.e. the dual string)
in $N=1, D=4$
\eqn\onet{\eqalign{ds^2&=-dt^2 + dx_1^2 + T_1 (dx_2^2
+ dx_3^2),\cr
T&=T_1 + iT_2=-{1\over 2\pi} \ln z.}}
Then it is easy to show that this configuration breaks
precisely half the spacetime supersymmetries of \ssnone\
by imposing two conditions on the components of $\epsilon$.
A quick check shows that the presence of additional nontrivial
$S$ and $T$ fields with the same analyticity behaviour
also lead to solutions of \ssnone, and this scenario
generalizes to multi-string solutions. The number, location and
winding numbers of the multi-string solitons is not relevant,
but the fact the fields have $\ln z$ or $\ln \bar z$ behaviour
is.
Provided the $S$ and various $T$ fields all have the same
analyticity (i.e. either all analytic or all anti-analytic) in
$z$,
then no new chirality choice is imposed by the addition of more
fields.
This can
be seen simply from the fact that the spin connection and
potential $G$ both scale logarithmically with the fields,
while $\delta \chi_{iL}$ is satisfied in the identical
manner for each $i$. Unlike the
$N=4$ case, the presence of these additional fields
produces no new conditions on $\epsilon$, as the
supersymmetry variations act on $\epsilon$
in precisely the same manner for all the fields. It follows
then that the $N=1$ analogs of these particular $N=4$ solutions
discussed
above break only half the spacetime supersymmetries in $N=1$,
and in some sense are realized naturally as stable
solitons only
in this context. When at least one of the fields, either $S$
or one of the $T$ fields, has a different analyticity behaviour
from the rest, opposite chirality conditions are imposed on
$\epsilon$, and no supersymmetries are preserved.

 A similar analysis can be done in the
$N=2$ case, at least for those solutions which can arise in
$N=2$.

\newsec{$S$, $T$ and $O(8,24;Z)$ Duality}
We now turn to
the issue of whether
$S$-duality and $T$-duality can be combined
in a larger duality group $O(8,24;Z)$
\refs{\dufldr\dufr\sentd{--}\hult}.
Their combination into $O(2,2;Z)$ was discussed
in \bakone.
In particular, in heterotic string theory
compactified on a seven-torus the $SL(2,Z)$ $S$-duality
should be combined with the $O(7,23;Z)$ $T$-duality into the
group
$O(8,24;Z)$. The existence of a Killing direction in all of the
solitons discussed in this paper means that they may all be
viewed as point-like solutions of three-dimensional heterotic
string theory. Thus from the viewpoint of three-dimensional
heterotic string theory $O(8,24;Z)$ appears as a duality group,
whereas from the point of view of four-dimensional heterotic
string
theory it appears as a solution-generating group.

Central to the issue of combining $S$ and $T$-duality is whether
there exists an $O(8,24;Z)$  transformation that
maps the fundamental $S$ string of \dabghr\ to the dual
$T$ string of \dufkexst.
In \sentd, Sen finds
a transformation that takes the
 fundamental string to a limit of
the stringy ``cosmic'' string solution
of \gresvy\ (with the slight subtlety that the two strings
point in different directions from the ten-dimensional
viewpoint).
We call this solution the $U$ string, since its
nontrivial field is (with a change of coordinates)
$U=g_{44}^{-1}(\sqrt{{\rm det} g_{mn}}
-ig_{45})$
with $m,n=4,5$.
This modulus
field transforms under $SL(2,Z)$ $U$-duality, just like $S$ and
$T$
do under $S$- and $T$-duality. It is related to the $T$ field
by an
$O(2,2;Z)$ transformation.
Therefore, it is no surprise that we can map the $S$ string to
the
$T$ string by an $O(8,24;Z)$ transformation.
The explicit
transformation doing this is
given as follows.
Following Sen's notation \sentd, ${\cal M}_T$, the
$32\times 32$ matrix that corresponds to the $T$ string, is
obtained from ${\cal M}_S$, the
$32\times 32$ matrix that corresponds to the $S$ string,
simply by
exchanging (rows and columns) $1$ with $10$, $2$ with $31$,
$3$ with $8$ and $9$ with $32$. The transformation matrix is
therefore, effectively, an  $O(4,4;Z)$ matrix.
It follows that
all three strings, fundamental, dual and cosmic are
related by $O(8,24;Z)$. As will be discussed in a subsequent
publication \drtwo, the $S$, $T$ and $U$ strings are related by
a
four-dimensional string/string/string triality.

Repeating Sen's and other arguments on three-dimensional
reduction for $N=2$ superstring theory, we can infer that
the larger duality group (or solution-generating
group) for $N=2$ containing $S$ and $T$-duality
is connected to a dual quaternionic manifold. In the case
of $n$ moduli, this group is $SO(n+2,4;Z)$
\refs{\cecfg,\ferquat}.

Of course, all the $N=4$ soliton
discussed in this paper (and many other solutions) are related
by generalized Geroch group transformations, because the
$D=4$ space-time admits two Killing directions. Therefore, the
theory is effectively two-dimensional and the equations
of motion have an affine $o(8,24)$ symmetry
\refs{\sengeroch,\dufr}.

\newsec{Conclusions}

The previously-known string soliton solutions, which break half
of the supersymmetries,
have played a crucial part in understanding
heterotic/Type IIA string/string duality. It is not yet clear
what part will be played by the new string soliton solutions
presented here which break more than half the supersymmetries.
These solutions
are in some sense realized naturally as stable
solitons only in the context of either $N=1$ or $N=2$
compactifications,
and
should lead to a better understanding of duality in
$N=1$ and $N=2$ compactifications and to
the construction of the Bogomol'nyi spectra of
these
theories. In these two cases, however, the situation is
complicated by
the absence of non-renormalization theorems present in the
$N=4$ case
which guarantee the
absence of quantum corrections. An exception
to this scenario arises for $N=2$ compactifications with
vanishing
$\beta$-function. The construction of these spectra remains a
problem
for future research.

In heterotic/Type IIA duality, we have $dH\neq 0$, but
$d\tilde{H}=0$. It is tempting to speculate that the dyonic
solution, for which both
$dH$ and $d\tilde{H}$ are non-zero,
will be important for the conjectured \dufm, but as yet little
explored, six-dimensional heterotic/heterotic duality.

\vskip1truecm

\noindent
{\bf Acknowledgements:}

\noindent
M. J. D. is grateful to the Director and the Staff of the Isaac
Newton Institute, where part of this work was completed.
R. R. K.
is grateful to Soo-Jong Rey for helpful discussions and to the
Princeton University Physics Department and Institute for
Advanced
Study at Princeton
for their hospitality and where part of this work was completed.

\vfil\eject
\listrefs
\bye